\documentclass[aps,pre,showpacs,noshowkeys,amsmath,amssymb,amsfonts,reprint]{revtex4-1}
\usepackage{graphicx}
\usepackage{bm}
\def\eq#1{{Eq.~\ref{#1}}}    \def\fig#1{{Fig.\ref{#1}}}
\def\be{\begin{equation}}   \def\ee{\end{equation}}
\usepackage[colorlinks=true,linkcolor=blue,citecolor=blue,urlcolor=blue]{hyperref}
\usepackage[all]{hypcap}
\begin{document}
\title{Model for probing membrane-cortex adhesion\\by micropipette aspiration and fluctuation spectroscopy}
\author{Ricard Alert}
\affiliation{Departament d'Estructura i Constituents de la Mat\`{e}ria, Universitat de Barcelona, Barcelona, Spain}
\author{Jaume Casademunt}
\affiliation{Departament d'Estructura i Constituents de la Mat\`{e}ria, Universitat de Barcelona, Barcelona, Spain}
\author{Jan Brugu\'{e}s}
\email{brugues@mpi-cbg.de}
\affiliation{Max Planck Institute of Molecular Cell Biology and Genetics, Max Planck Institute for Physics of Complex Systems, Dresden, Germany}
\author{Pierre Sens}
\email{pierre.sens@curie.fr}
\affiliation{Laboratoire Gulliver, CNRS-ESPCI Paris Tech, UMR 7083, Paris, France}
\date{\today}
\begin{abstract}
We propose a model for membrane-cortex adhesion which couples membrane deformations, hydrodynamics and kinetics of membrane-cortex ligands. In its simplest form, the model gives explicit predictions for the critical pressure for membrane detachment and for the value of adhesion energy. We show that these quantities exhibit a significant dependence on the active acto-myosin stresses. The model provides a simple framework to access quantitative information on cortical activity by means of micropipette experiments. We also extend the model to incorporate fluctuations and show that detailed information on the stability of membrane-cortex coupling can be obtained by a combination of micropipette aspiration and fluctuation spectroscopy measurements.
\end{abstract}
\maketitle
\section{Introduction}
In many cells, a thin layer of cytoskeleton called cortex underlies the plasma membrane. While the cellular membrane serves as a barrier for the cell and a mechanism to communicate with the extracellular media, the cortex, made mostly of actin cross-linked filaments and myosin II, provides rigidity and allows for active remodelling of the cell boundaries, essential for instance for cell motility. The control of membrane-cortex adhesion is crucial to many cellular processes. Indeed, membrane-cortex detachment and the formation of cellular blebs, spherical protrusions of the unbound plasma membrane, is often a sign of apoptosis \cite{Coleman2001,Vermeulen2005}. Membrane blebbing is also used for motility by several cell types, including amoebae and possibly cancer cells \cite{Blaser2006,Yoshida2006,Fackler2008,Charras2008}.

It is acknowledged that membrane-cortex adhesion is obtained via specific interactions between large numbers of ligand and receptor molecules \cite{Sheetz2001}, such as Talin \cite{Tsujioka2012} and ERM (Ezrin/Radixin/Moesin) Proteins \cite{Tsukita1999}. Spontaneous membrane detachment, also known as blebbing, has been associated to myosin activity within the cortex \cite{Charras2008a,Tinevez2009}. Externally induced perturbations using micropipette aspiration or osmotic shocks, show that a sufficiently large drop of external pressure can induce membrane detachment \cite{Rentsch2000}. Consequently the links between the membrane and cortex are constantly under stress, which origin is ultimately related with acto-myosin cortical tension and osmotic pressure. 

In this article, we present a model for adhesion based on the kinetics of the membrane-cortex ligands \cite{Seifert2000,Erdmann2004,Erdmann2008,Brugues2010}. We describe the stability of adhesion by coupling the kinetics of the ligands to the stress exerted on them and to physical properties of the membrane. In its simplest form, the model establishes the mechanical equilibrium of the cell considering both the pressure drop across the membrane and the pre-stressed state of the cortex, and predicts the outcome of a micropipette aspiration experiment in terms of physical parameters. This predictions are then compared to experiments from the literature. We also discuss extensions of the model to include spatial modulations of the membrane and different scenarios of hydrodynamic interactions, depending on the porosity of the cortex and its actual distance to the membrane. In particular, we obtain analytical expressions for the structure factor and fluctuation spectrum of the membrane in certain limits, and show how these results may be used to obtain additional information on the density of ligands by means of fluctuation spectroscopy experiments on eukaryotic cells.

\section{Model for membrane-cortex adhesion}
The adhesion of a flexible membrane on a substrate by means of discrete linkers has been extensively studied in the past \cite{Rozycki2006,Reister-Gottfried2008,Krobath2009,Weikl2009,Reister2011,Hu2013}, mostly using computer simulations. It is a highly non-trivial problem due to the multiplicity of energy scales (membrane rigidity and tension, linker stiffness and binding energy) and time scales (membrane and cytosol fluidity, linker's diffusion and binding kinetics). In particular, the role of fluctuations on the unbinding transition of a membrane possessing meta-stable bound and unbound states has been characterised numerically \cite{Reister-Gottfried2008}, but the unbinding of a membrane subjected to a constant pressure has, to our knowledge, not been systematically investigated. Our primary goal here is to assess the role of cortical prestress on membrane-cortex detachment.

\begin{figure}[tbp]
\begin{center}
\includegraphics[width=7.5cm]{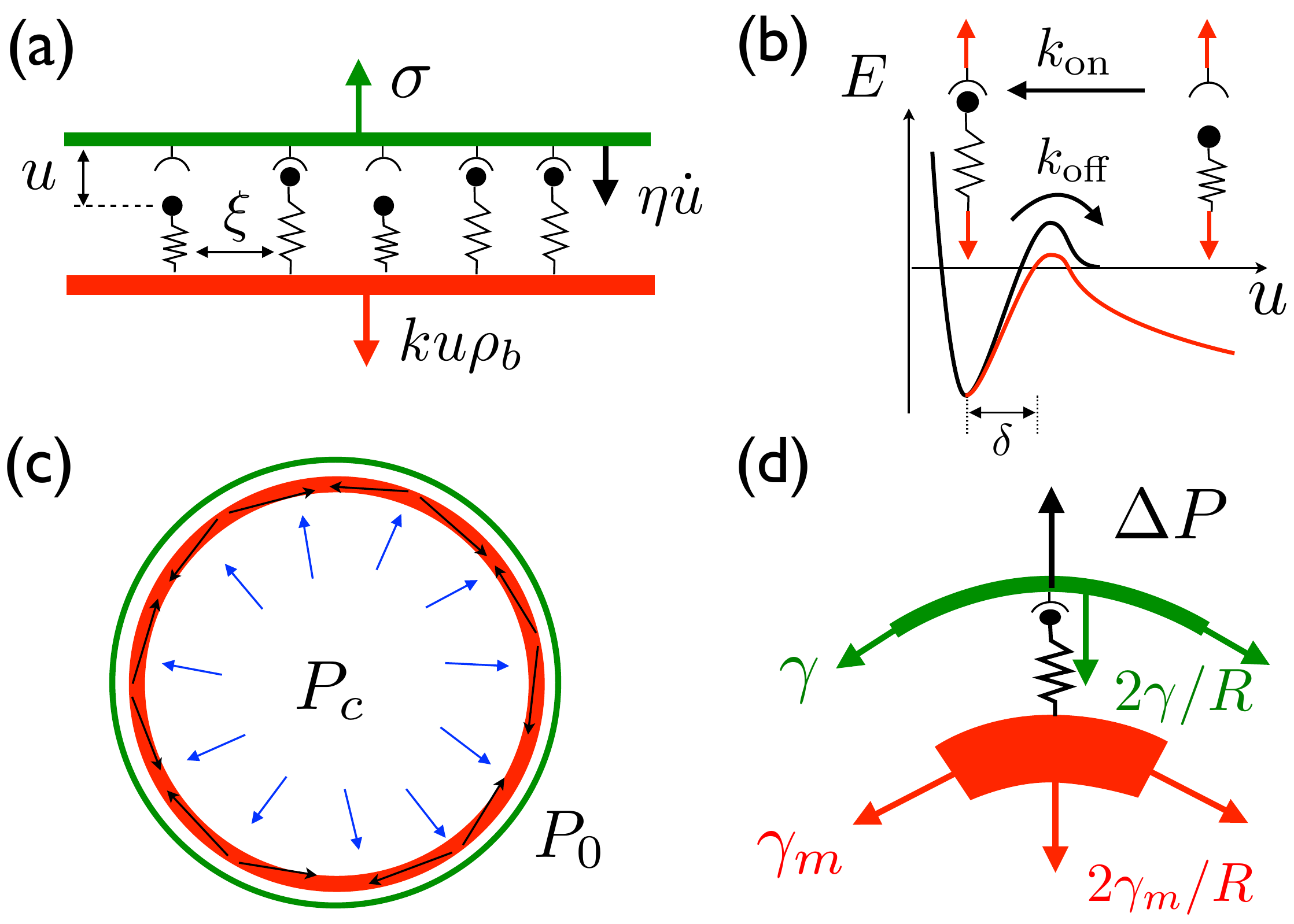}
\caption{Sketch of the system. (a) The ligands are modeled as springs that link the cortex (red) and the membrane (green). (b) Kinetic rates $k_{\rm on}$ and $k_{\rm off}$ of the ligands. $k_{\rm off}$ depends on the load \cite{Evans2001}. (c) Forces involved in the cell at steady state: internal pressure, $P_c$, and external pressure, $P_0$, exert a normal force on the membrane and cortex, which is compensated by the membrane and cortex tension. (d) The normal projection of the acto-myosin tension in the cortex is transmitted to the membrane through proteins that link the cortex and the membrane.}
\label{laplace_law}
\end{center}
\end{figure}

To this aim, we first adopt a highly simplified model, where we assume a nearly planar membrane subject to a normal external stress $\sigma$ and attached to the cortex by a density of linkers $\rho_b$, which is necessarily smaller than a maximal value $\rho_0$ (\fig{laplace_law}). The cortex is assumed to be flat and immobile, so that the model is only valid at length scales below the correlation length for cortex undulations. For a constant normal stress $\sigma$, an equilibrium state may exist with a planar membrane at position $u$ where a uniform density $\rho_b$ of bound spring-like linkers with elastic constant $k$ balances the external force. In order to find the conditions for the existence and stability of such an equilibrium state we may write dynamical equations assuming spatial uniformity, where $u$ and $\rho_b$ are only time-dependent:
\begin{align}
\eta \frac{du}{dt}& = \sigma - k u \rho_b, \label{bond_eq_1}\\
\frac{d\rho_b}{dt} &= k_{\rm on}\left[\rho_0-\rho_b\right] - k_{\rm off}(u) \rho_b, \label{bond_eq_2}
\end{align}
where $\eta$ is an effective viscosity per unit length, and $u=0$ corresponds to the position for which the bound linkers are not stretched. For small membrane displacements, the relevant contribution to dissipation is due to cytosol flow through the cortex meshwork, and the effective parameter  $\eta$ can be estimated as $\eta \sim \eta_c h/\xi^2$ (see section 1 in the Supporting Material for details), where $\xi\sim 30$ nm is the scale of the cortex mesh size \cite{Bovellan2014}, $h\sim 500$ nm is the thickness of the cortex, and $\eta_c\sim3\times 10^{-3}-2\times 10^{-1}$ Pa s is the cytosol viscosity \cite{Charras2008a}.

The linker kinetics is defined by the attachment and detachment rates $k_{\rm on}$ and $k_{\rm off}$ (\fig{laplace_law}), and is assumed to be much faster than the typical time scale of membrane shape relaxation. The force-dependent kinetics of the linkers then imposes a strong nonlinear coupling between the kinetics and the position of the membrane. The detachment rate is assumed to follow a Kramers-like kinetics \cite{Kramers1940} appropriate of thermally induced processes:
\begin{equation}
k_{\rm off}(u)=k_{\rm off}^0 e^{ku\delta/(k_B T)},
\end{equation}
where $\delta$ is a characteristic bond length in the nanometric scale \cite{Evans2001}. For simplicity, we assume linker attachment to be an active process occurring at a constant rate $k_{\rm on}$. Therefore, detailed balance is not obeyed, as previously considered in membrane adhesion problems \cite{Rozycki2006}. This assumption allows to disregard membrane fluctuations between attachment points and yield a simple analytical form for the unbinding transition. However, it  does not capture binding cooperativity occuring due to the smoothing of membrane fluctuations near attachement points  \cite{Reister-Gottfried2008,Krobath2009,Weikl2009,Reister2011,Hu2013}.

Two relevant dimensionless quantities characterize the mechanics of the linkers: the kinetic ratio, $\chi$, and the ratio of the force on the membrane to an intrinsic force scale of the linkers, $\alpha$, with
\begin{equation}
\chi\equiv \frac{k_{\rm off}^0}{k_{\rm on}} \qquad{\rm and}\qquad \alpha\equiv \frac{\sigma \delta}{\rho_0 k_BT}.
\label{alpha}
\end{equation}
Equilibrium solutions to \eq{bond_eq_1}-\eq{bond_eq_2} exist only for $\alpha < \alpha^*$ where the latter is defined by:
\begin{equation}
\alpha^* e^{1+\alpha^*}=\chi^{-1}.
\label{alphastar}
\end{equation}
Taking $\chi\sim 10^{-3}$ \cite{Rognoni2012} and $\delta\sim 1$ nm, the critical force per link is $\sigma^*/\rho_0\sim 18$ pN, corresponding to $\sim 4.5$ times the thermal force per link $k_BT/\delta$. This fixes the condition for the detachment of the membrane from the cortex, which occurs for stresses that surpass the critical stress $\sigma^*=\rho_0\alpha^*(\chi)k_B T/\delta$.

The adhesion energy $w$ per unit area may be defined as the work necessary to bring the stress of the linkers from its rest value to the critical value for detachment in a quasi-static fashion, that is,
\begin{equation}\label{work}
w\left(u_{\rm eq}\right) = \int_{u_{\rm eq}}^{u^*} \sigma (u) du =  \rho_0 k \int_{u_{\rm eq}}^{u^*} \frac{u}{1+\chi e^{ku\delta/(k_B T)}} du,
\end{equation}
where $\sigma(u)$ is the equilibrium stress for each $u$. Note that the adhesion energy depends on the actual state of the cell $u_{\rm eq}$, which is generically unknown and incorporates the pre-stress state of the cell due to cortical tension.

Within our simplified model, the average density of bound linkers $\rho_{b,\text{eq}}$, the critical stress $\sigma^*$, and the adhesion energy $w$ all scale linearly with the density of available linkers $\rho_0$. This scaling results from our assumption of a constant binding rate. A different scaling is expected if the binding rate depends on the average position and fluctuations of the free membrane between anchoring points. If the on-rate obeys detailed balance, one expects $\rho_{b,\text{eq}}\sim\rho_0^2$ in the absence of a pressure difference \cite{Krobath2009,Weikl2009}. As discussed in the following sections, the results of micropipette experiments are consistent with a linear scaling $\sigma^*\sim \rho_0$.

\section{Results and Discussion}
The simplified stochastic model of adhesion outlined in the previous section is used below to analyse two different kinds of experiments that can probe membrane-cortex interaction. First, we analyse micropipette experiments where the critical suction pressure required to unbind the cell membrane from the cortex was measured in different cellular contexts, where the density of adhesion molecules and of cortical motors have been altered. Second, we derive the effect of membrane-cortex interaction on the membrane fluctuation spectrum. There is as of yet no experimental data that can be directly confronted to the latter derivation. We hope that the present paper will foster experimental spectroscopy studies that will couple membrane fluctuation analysis with cell micromanipulation, along the lines described in Sec.\ref{spectro} below.

\subsection{Mechanical equilibrium of the cell}
Force balance at the membrane involves the difference of pressure across the membrane, $\Delta P$, and the normal projection of the cortex and membrane tension, $\gamma_m$ and $\gamma$, respectively:
$\Delta P=2 \left( \gamma_m + \gamma\right)/R$,
where $R$ is the radius of the cell, assumed spherical. At equilibrium, the links sustain the stress needed to maintain the cortex and the membrane adhered, $\sigma_{\rm eq}=2\gamma_m/R$, which accounts for the difference between the pressure and the membrane tension stresses, $\Delta P -  2\gamma/R$. Whenever the equilibrium stress exceeds the critical value $\sigma^*$, we expect the cell membrane to detach spontaneously.
 
Micropipette aspiration \cite{Dai1999a,Rentsch2000,Merkel2000,Brugues2010,Campillo2012}, amongst other techniques \cite{Dai1999,Tinevez2009,Diz-Munoz2010}, allows to apply pressure perturbations of controlled intensity and area. Pressure perturbations can be supplemented with perturbations on relevant cell parameters such as myosin activity and link or cortex density, by genetics \cite{Dai1999a,Merkel2000,Campillo2012} or direct drug treatment \cite{Charras2008a,Tinevez2009,Diz-Munoz2010}. Tether pulling experiments have also been used to probe membrane-cortex adhesion \cite{Borghi2007}, but their interpretation is rather non-trivial \cite{Schumacher2009}. In the following, we restrict ourselves to a quantitative interpretation of micropipette aspiration experiments.

\subsection{Micropipette aspiration}
During a micropipette experiment, a pressure drop is applied on a small region of the membrane defined by the micropipette radius $R_p$. A new equilibrium state in the micropipette requires an increase of the stress exerted on the links with respect to $\sigma_{\rm eq}$:
\begin{equation}\label{force_pipette_myosin}
\sigma= \Delta P_p - 2\gamma\left(\frac{1}{R_p} - \frac{1}{R}\right)+  2\frac{\gamma_m}{R},
\end{equation}
where $\Delta P_p\equiv P_0-P_p$ is the difference between the extracellular media and the aspiration pressure, and $R$ is the radius of the cell after deformation. Characteristic bounds for membrane tension $\gamma\lesssim 10^{-4}$ N/m and radius of cell $R\sim 10$ $\mu$m and pipette $R_p\sim 5$ $\mu$m allow membrane tension to compensate for a pressure of about $\sim 20$ Pa, which is small compared to the range of experimental pressures $\sim100-1000$ Pa. As a consequence, we will neglect the membrane tension contribution in the following. The last term in the right hand side accounts for the cortical stress, or pre-stressed state of the cell $\sigma_{\rm eq}$. In general, force balance does not need to be satisfied and the cell will eventually be entirely sucked inside the pipette if the suction pressure $\Delta P_p$ is too large \cite{Brugues2010}. Here we focus on the case where the cortex is able in principle to compensate for the pipette pressure.

\begin{figure}[tbp]
\begin{center}
\includegraphics[width=7.5cm]{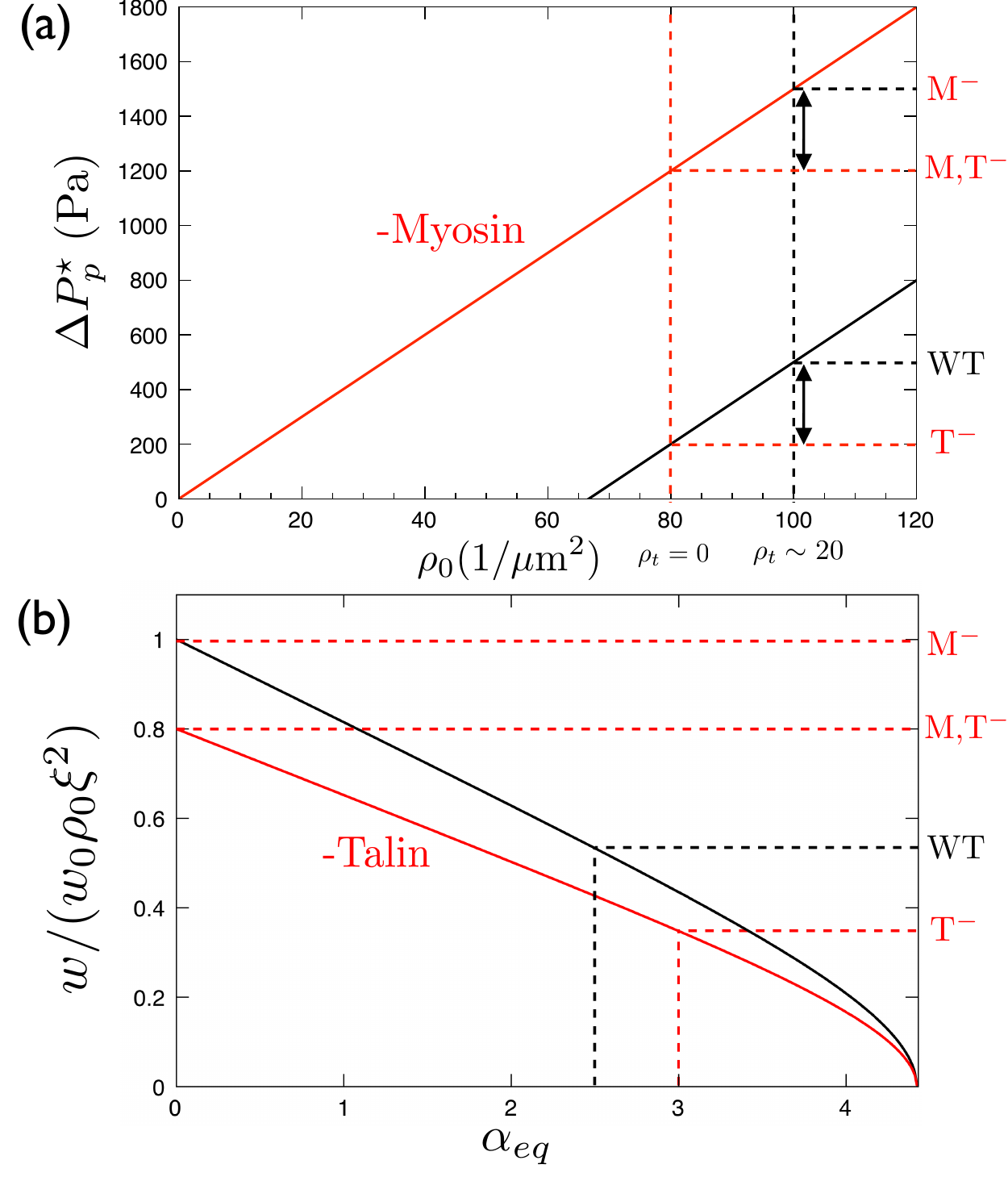}
\caption{Theoretical predictions for the critical aspiration pressure and adhesion energy in a micropipette experiment. (a) Critical pressure as a function of the density of linkers $\rho_0$ according to \eq{pipette_vs_tension}. Solid black and red lines correspond to cells with and without myosin II, respectively. The horizontal dashed lines are the experimentally measured value of the critical detachment pressure \cite{Merkel2000} for wild type cells (WT), mutants lacking myosin ($M^-$), mutants lacking talin ($T^-$) and double mutants ($M,T^-$). The slope and height of the two theoretical curves are entirely determined by these critical pressures (see text). (b) Effective adhesion energy as a function of the equilibrium cortical tension in the cell according to \eq{adhesion_energy_2}. Solid black and red lines correspond to cells with and without talin, respectively.}
\label{pressure_merkel_theory}
\end{center}
\end{figure}

Using our previous analysis for the membrane-cortex adhesion, we can relate the critical stress for the links, $\sigma^*$, with the critical aspiration pressure needed to unbind the membrane via \eq{force_pipette_myosin}:
\begin{equation}\label{pipette_vs_tension}
\Delta P_p^{*} = \rho_0\alpha^* \frac{k_B T}{\delta}-2\frac{\gamma_m}{R}.
\end{equation}
The critical aspiration pressure has two contributions: the pressure needed to detach a certain number of relaxed links, given by the density of ligands and the critical force per link (first term), and  the contribution from the presence of acto-myosin tension in the cortex which sets a non-zero stress on the links at equilibrium, hence reducing the amount of pressure needed to reach the critical stress (second term, \fig{pressure_merkel_theory}a).

As in determining the critical aspiration pressure, we find that the adhesion energy per unit area measured when detaching the membrane (\eq{work}) depends on the level of cortical rest tension, $\sigma_{\rm eq}=2\gamma_m/R$, which ultimately determines the effective number of ligands to be broken:
\begin{equation}\label{adhesion_energy_2}
w = w_0 \bar{\rho}_0 \int_{z_{\rm eq}}^{z^*} \frac{z}{1+\chi e^z} dz.
\end{equation}
Here, $w_0 \equiv (k_B T/\delta)^2 / (k\xi^2)$ is an upper bound for the adhesion energy, that corresponds to non pre-stressed ligands, and for clarity we have used rescaled quantities for the stretching, $z\equiv u/u_0$ with $u_0\equiv k_BT/\left(k\delta\right)$, and ligand density $\bar{\rho}_0\equiv\rho_0 \xi^2$. The adhesion energy per unit area depends linearly on the saturation density of links, $w \sim w_0 \bar{\rho}_0$, but contains a correction factor that includes the pre-stressed state of the cell. In the presence of cortical tension in the cell, there is both a reduction of the number of effective bound links, and an increase of stress per link. Consequently, close to the unbinding transition, the adhesion energy is reduced in a strongly non-linear way by increasing the cortex prestress (\fig{pressure_merkel_theory}b).

\subsection{Discussion of micropipette experiments}
Our model allows to directly relate the critical perturbation pressure needed to detach the membrane from the cortex to two physiologically relevant quantities: the density of membrane-cortex ligands, and the myosin-driven cortical tension (\eq{pipette_vs_tension}). This relationship provides not only a rationale explanation for the membrane unbinding for a variety of cell phenotypes where either the density of ligands or myosin activity is altered, but also a method to directly probe cortex activity by measuring the critical pressure needed to unbind the membrane.

We refer to previous experimental results concerning the abrupt unbinding induced by micropipette suction to assess the validity of our model \cite{Rentsch2000,Merkel2000}. In order to test the relationship between critical pressure, ligand density and cortical tension, we would ideally need to measure the critical pressure for cells whose phenotype has been quantitatively altered. Merkel et al. \cite{Merkel2000} considered four phenotypes of the amoebae {\it Dictyostelium}: wild type, myosin inhibited, talin inhibited (a membrane-cortex linker), and double mutants. These four phenotypes are sufficient to qualitatively test our model and obtain values for all the relevant parameters.

Mutations that perturbed ligand density and cortex activity should be independent within our model. Accordingly, the difference of unbinding pressure for two values of ligand density must be the same independently of the value of cortical activity (\fig{pressure_merkel_theory}a). In Merkel et al. \cite{Merkel2000}, the decrease of critical pressure between the wild type and talin inhibited amoebae is comparable to the corresponding decrease between the myosin inhibited and double inhibited mutants ($\sim 150-200$ Pa and $\sim 150-500$ Pa respectively), even though the actual values of the cortical tension with and without myosin differ by a factor of 5 due to cortical prestress. This suggests that the critical pressure scales linearly with the density of available bounds: $\Delta P^*_p\sim\rho_0$, as predicted by our simple model (\eq{pipette_vs_tension}). Comparing the critical pressures in both wild type and myosin-null cells for a fixed link density (Fig. 3b-4b in \cite{Merkel2000}), we can estimate the myosin-driven cortical stress in the wild type amoeba, $\gamma_m = (\Delta P_p^{*M^-} - \Delta P_p^*)R/2 \sim 5 \times 10^{-3}$ N/m. This is at least two orders of magnitude higher than the typical membrane tension of a vesicle, $\gamma$, and contributes to the $60 \%$ of the $\sim 1600$ Pa needed to unbind the membrane. This estimate of the cortical tension agrees well with direct experimental measurements in \textit{Dictyostelium} \cite{Dai1999a}. Finally, introducing the obtained value of $\gamma_m$ into the rest stress $\sigma_{\rm eq}=2\gamma_m/R$, and using the stationary state solution of Eqs.~\ref{bond_eq_1}-\ref{bond_eq_2}, $z_{\rm eq}=\alpha_{\rm eq}\left(1+\chi e^{z_{\rm eq}}\right)$, the equilibrium stretching of the linkers can be found, $u_{\rm eq}\sim 100$ nm, as well as that roughly all the linkers are connected in equilibrium conditions for the wild-type cells, $\rho_{b,\rm eq}/\rho_0=\alpha_{\rm eq}/z_{\rm eq}\sim 1$.

For myosin-inhibited amoebae, the micropipette pressure is directly related to the available density of links (\eq{pipette_vs_tension}). Using the results from \cite{Merkel2000}, we can estimate the relative concentration of talin with respect to the saturation link concentration: $\rho_t/\rho_0 =(\Delta P_p^{*M^-}- \Delta P_p^{*M,T^-})/\Delta P_p^{*M^-}\sim 10-30\%$. Assuming the saturation density to be $\rho_0 \sim 100$ links/$\mu$m$^2$, talin density should be roughly $\rho_t\sim 20$ links/$\mu$m$^2$. The asymmetric distribution of this small density of talin links seems to be enough to drive direct motion in amoebae \cite{Merkel2000}. Similar observations are reported for zebrafish cells \cite{Diz-Munoz2010}. For completeness, assuming a ligand length $\delta\sim 1$ nm, we find $\alpha^*=\Delta P_p^{*M^-}\delta/\left(\rho_0k_BT\right)\sim 4$, and the critical force per link $\sigma^*/\rho_0\sim 16$ pN is four times the thermal force of the link $k_BT/\delta$, which is close to our initial estimate ($\sim 18$ pN). This quantity is independent of the cell phenotype and only depends on the kinetic rate ratio $\chi$. In fact, from the experimental estimate of $\alpha^*$ we can derive the kinetic ratio of on and off rates of the membrane-cortex linkers, $\chi\sim 10^{-3}$, in agreement with \cite{Rognoni2012}. Moreover, using the stationary solution of our model, a critical stretching $u^*\sim 200$ nm and a critical fraction of bound linkers $\rho_b^*/\rho_0\sim 0.9$ are found. Our results show that the rest stress $\sigma_{\rm eq}=2\gamma_m/R$ is about $60\%$ of the critical unbinding value $\sigma^*$ for wild-type cells, while it is around $75\%$ in talin-null cells. This is consistent with the observation that spontaneous blebbing of migratory {\it Dictyostelium} is more frequent for talin-null mutants than for wild-type cells \cite{Zatulovskiy2014}.

\begin{table}[t]
\begin{center}
\begin{tabular}{clc}\hline
Symbol&Description&Estimate (Ref.)\\\hline
$\xi$&cortex mesh size&$30$ nm \cite{Bovellan2014}\\
$h$&cortex thickness&$500$ nm \cite{Charras2008a}\\
$\eta_c$&cytosol viscosity&$10^{-2}$ Pa s \cite{Charras2008a}\\
$k_{\text{on}}$&linker attachment rate&$10^4$ s$^{-1}$ \cite{Rognoni2012}\\
$k_{\text{off}}^0$&free linker detachment rate&$10$ s$^{-1}$ \cite{Rognoni2012}\\
$\delta$&linker bond length&$1$ nm \cite{Evans2001}\\
$k$&linker stiffness&$10^{-4}$ N/m (text)\\
$\rho_0$&density of available linkers&$10^{14}$ m$^{-2}$ (text)\\
$R$&cell radius&$10$ $\mu$m \cite{Merkel2000}\\
$\gamma$&membrane surface tension&$5\times 10^{-5}$ N/m \cite{Tinevez2009}\\
$\kappa$&membrane bending ridigity&$10^{-19}$ J \cite{Dai1999}\\
$\gamma_m$&cortical tension&$5\times 10^{-3}$ N/m (this work)\\\hline
\end{tabular}
\end{center}
\caption{Estimates for model parameters.} \label{table-parameter}
\end{table}

Finally, our model gives a prediction for the adhesion energy as a function of the ligand density and cortical activity (\eq{adhesion_energy_2}). In the case of the four phenotypes discussed above, the maximum adhesion energy is $w_0\rho_0\xi^2\sim 2\times 10^{-5}$ J/m$^2$, and corresponds to the mutant lacking myosin (a non pre-stressed cell, $\alpha_{\rm eq}=0$). For a mutant lacking Talin and myosin II, the adhesion energy is reduced by $10-30\%$ due to the decrease in $\rho_0$. For a wild type cell and a mutant lacking talin the adhesion energies are further reduced, by a $50\%$ and $65\%$ respectively, due to cortical pre-stress (\fig{pressure_merkel_theory}b). The dramatic increase in the adhesion energy for a cell lacking myosin activity, which can be of the order of $200\%$, illustrates the importance of cortex activity in the cell in determining the experimental measurements of adhesion energy and detachment pressures. Table \ref{table-parameter} recapitulates the numerical values used for the parameters of the model. These parameters may vary significantly depending on cell lines and experimental conditions, so this choice is somewhat arbitrary. However, we emphasize that both the cortical tension $\gamma_m$ and the fraction of bond that are associated with Talin $\rho_t/\rho_0$ do not depend on this choice and can be directly determined by confronting \eq{pipette_vs_tension} with the experimental results.


\subsection{Membrane undulations}
The model for membrane-cortex adhesion discussed so far considers a flat membrane, disregarding possible membrane undulations. In this section, we address the linear dynamics of long-wavelength perturbations around the flat membrane state:
\begin{align}
u\left(\vec{x},t\right)&=u_{\rm eq}+\delta u\left(\vec{x},t\right),\\
\rho_b\left(\vec{x},t\right)&=\rho_{b,\rm eq}+\delta\rho_b\left(\vec{x},t\right).
\end{align}
The coarse-grained interface hamiltonian includes the elastic energy of bound linkers and contributions from the membrane bending rigidity and tension \cite{Boal2002}:
\begin{multline} \label{eq Hamiltonian}
\mathcal{H}=\int_S\left[\frac{\kappa}{2}\left[\nabla^2 u\left(\vec{x}\right)\right]^2+\frac{\gamma}{2}\left[\vec{\nabla}u\left(\vec{x}\right)\right]^2\right.\\
\left.+\frac{k}{2}\rho_b\left(\vec{x}\right)u^2\left(\vec{x}\right)-\sigma u\left(\vec{x}\right)\right]d^2\vec{x},
\end{multline}
where $\kappa$ is the bending modulus and where $\sigma=\rho_{b,\rm eq}ku_{\rm eq}$. As before, the restoring elastic forces exerted by the linkers is treated within a continuous approximation, and membrane fluctuations between bound linkers are not accounted for. This description is appropriate for length scales larger that the average spacing between linkers $\rho_0^{-1/2}\sim 100$ nm, and the present analysis is only valid for length scales larger than this cutoff. 

Membrane deformations induce Stokes flows in the surrounding fluid. These flows mediate long-range hydrodynamic interactions in the membrane, leading to a non-local membrane dynamics that is better treated in Fourier space. The full dynamical problem requires a proper treatment of cytosol permeation through the porous cortex and the (less) porous lipid membrane at all length scales \cite{Gov2004a,Strychalski2013}. For simplicity, we restrict ourselves to a simplified treatment, where cytosol permeation through the cortex is only included for the lowest Fourier mode $q=0$. The other modes are treated below neglecting the effect of the cortex on hydrodynamics, as is appropriate for sufficiently large membrane-cortex distances and/or large cortex mesh size. The effect of finite cortex permeation is studied in section 4 in the Supporting Material. Using standard results of membrane hydrodynamics \cite{Seifert1997} together with \eq{eq Hamiltonian}, the dynamics of long-wavelength membrane deformations read
\begin{align}
\partial_t\delta\tilde{u}_{\vec{0}}&  =-\frac{1}{\eta}\left[\rho_{b,\rm eq}k\delta\tilde{u}_{\vec{0}}+\frac{\sigma}{\rho_{b,\rm eq}}\delta\tilde{\rho}_{b,\vec{0}}\right],\label{eq u0}\\
\partial_t\delta\tilde{u}_{\vec{q}}&  =-\frac{1}{4\eta_c q}\left[\left(\kappa q^4+\gamma q^2+\rho_{b,\rm eq}k\right)\delta\tilde{u}_{\vec{q}}+u_{\textrm{eq}}k\delta\tilde{\rho}_{b,\vec{q}}\right], \label{eq uq}
\end{align}
where $\vec{q}$ is the wave-vector. Within our approximation, the  relaxation dynamics of the mode $q=0$, \eq{eq u0}, is decoupled from the other modes, \eq{eq uq}, at the linear level of perturbations. \eq{eq u0} can be seen as a linearized version of \eq{bond_eq_2} when transformed back to real space.

In turn, the dynamics of the long-wavelength perturbations of the density of bonds reads
\begin{multline} \label{eq linkers}
\partial_t\delta\rho_b\left(\vec{x}\right)=-\frac{k\delta}{k_BT}k_{\rm off}^0e^{ku_{\rm eq}\delta/(k_BT)}\rho_{b,\rm eq}\delta u\left(\vec{x}\right)\\
-\left[k_{\rm on}+k_{\rm off}^0e^{ku_{\rm eq}\delta/(k_BT)}\right]\delta\rho_b\left(\vec{x}\right).
\end{multline}
\eq{eq u0}-\eq{eq linkers} completely specify the dynamics of linear perturbations around the flat membrane state, both for the membrane displacement $u$ and the density of bonds $\rho_b$. However, in the limit of long wavelengths, membrane deformations proceed much slower than linker kinetics. In general, membrane dynamics is slower than linkers kinetics at length scales above a crossover wavelength $\lambda_{\text{cross}}$, that is determined from an analysis of the eigenvalues and eigenvectors of the dynamical system \eq{eq uq}-\eq{eq linkers}. With the parameters given in Table \ref{table-parameter}, this crossover occurs in the bending-dominated regime, for which $\lambda_{\rm cross}\simeq 2\pi (\kappa/(4\eta_c k_{\rm on}))^{1/3} \sim 0.4$ $\mu$m. For larger length scales, the kinetics of the linkers \eq{eq linkers} is always essentially equilibrated and an adiabatic approximation may be used. The system can then be described in terms of only the slow variable $\delta u$:
\begin{equation} \label{eq adiabatic}
\partial_t\delta\tilde{u}_{\vec{q}}=-\frac{\kappa q^4+\gamma q^2+\rho_{b,\rm eq}k}{4\eta_c q}\delta\tilde{u}_{\vec{q}}.
\end{equation}

Under the adiabatic approximation, the dispersion relation of membrane dynamics $\omega\left(q\right)=-\left(\kappa q^4+\gamma q^2+\rho_{b,\rm eq}k\right)/\left(4\eta_c q\right)$ features a maximum due to membrane-cortex adhesion (see section 2.1 in the Supporting Material for details). This maximum naturally defines a correlation length for shape fluctuations, $\lambda_c$, below which the membrane can be seen as essentially rigid. This correlation length depends on a combination of both mechanical properties of the membrane and of the linkers:
\begin{equation} \label{correlation}
\lambda_c=2\pi\left[\frac{6\kappa/\gamma}{\left(1+12\kappa\rho_{b,\rm eq}k/\gamma^2\right)^{1/2}-1}\right]^{1/2}.
\end{equation}
With the values given in Table \ref{table-parameter}, we find $\lambda_c\sim 0.6$ $\mu$m for an unperturbed cell ($\rho_{b,\rm eq}\simeq\rho_0$). This value is larger than both the crossover wavelength of the free membrane undulations, $\lambda=2\pi\sqrt{\kappa/\gamma}\sim 0.3$ $\mu$m, and the spacing between linkers, $\rho_0^{-1/2}\sim 0.1$ $\mu$m. The computed correlation length is slightly smaller than the pipette radius, so the approximation of a rigid membrane is only marginally valid in that case. However, it becomes more accurate near the unbinding transition since the correlation length $\lambda_c$ increases with decreasing density of bonds $\rho_b$ (see section 3 of the Supporting Material for details). In the general case, including all hydrodynamic effects of the cortex, the value of $\lambda_c$ may differ from \eq{correlation} or, for low cortex porosity and short membrane-cortex distances, it may not even be well defined (see section 4 in the Supporting Material for details).

Finally, at the mean-field level, the critical stress $\sigma^*$ at which the membrane detaches from the cortex is not affected by membrane undulations since the $q=0$ mode is the first one to become unstable in the framework of \eq{eq u0}-\eq{eq linkers}. Fluctuations of the membrane shape may however create regions of locally low linker density and high linker stress, thereby widening the unbiding transition boundary.

\subsection{Fluctuation spectroscopy}\label{spectro}
The formulation of an adhesion model accounting for membrane undulations provides an appropriate framework to extract additional information about membrane-cortex adhesion from the statistics of membrane fluctuations. For instance, applying the energy equipartition theorem to \eq{eq Hamiltonian} one obtains, under the adiabatic approximation, a membrane structure factor
\begin{equation} \label{eq spectrum}
S\left(q\right)=\frac{k_BT}{\kappa q^4+\gamma q^2+\rho_{b,\rm eq}k},
\end{equation}
where $\rho_{b,\rm eq}$ is the equilibrium value of the density of bound linkers (see section 2.2 of the Supporting Material for details). This result is consistent with the situation of a membrane confined into an harmonic potential \cite{Gov2003,Fournier2004,Merath2006a}. Here, the confinement contribution explicitly arises from the attachment kinetics of the linkers via the adiabatic approximation. This fact allows to experimentally determine the density of bound linkers, $\rho_{b,\rm eq}$, from measurements of the static structure factor of the cell membrane \cite{Popescu2006}. Specifically, the long-wavelength limit $q\rightarrow 0$ needs to be measured in fluctuation microscopy experiments in order to determine $\rho_{b,\rm eq}$ from \eq{eq spectrum}. Transforming \eq{eq spectrum} to real space, the mean-square amplitude of membrane undulations reads (see section 2.3 of the Supporting Material for details):
\be
\sqrt{\left\langle\delta u^2\right\rangle}\simeq \sqrt{\frac{k_BT}{8\sqrt{\kappa\rho_{b,\rm eq}k}}}\sim 4\, {\rm nm}.
\ee

\begin{figure}[!htbp]
\begin{center}
\includegraphics[width=7.5cm]{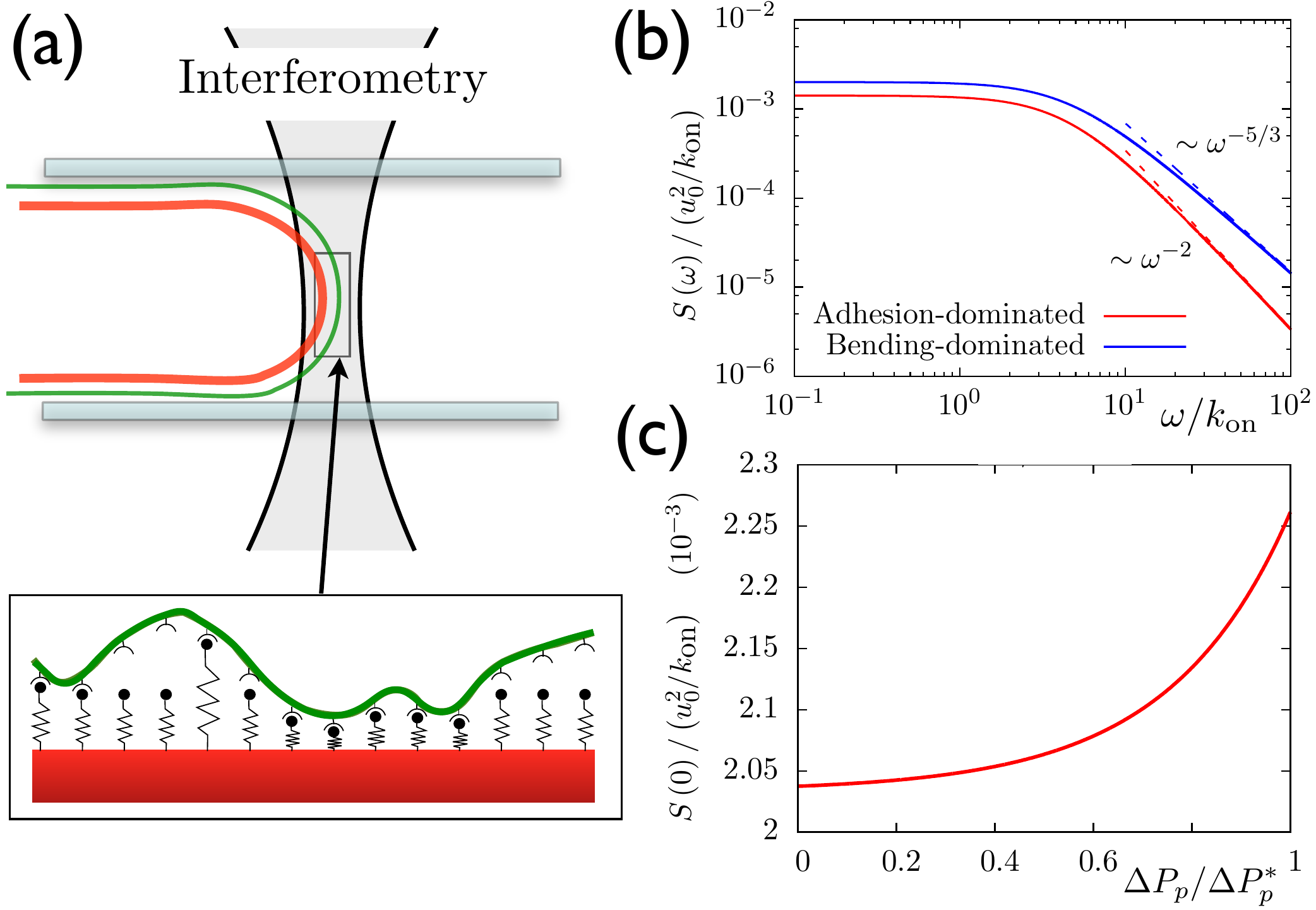}
\caption{Density of membrane-cortex bonds from fluctuation spectroscopy experiments. (a) Illustration of a combined spectroscopy and micropipette experiment that could probe the density of membrane-cortex bonds. (b) Power spectral density calculated from \eq{eq_psd} in the limit of vanishing surface tension ($\gamma=0$), both for adhesion-dominated and bending-dominated membrane fluctuations. The known high-frequency limits are indicated in dashed lines. The rescaling length $u_0$ is defined as $u_0\equiv k_BT/(k\delta)$. Parameters are taken from Table \ref{table-parameter}, with $\rho_{b,\text{eq}}=\rho_0$, and the power spectrum is integrated from $q_{\text{min}}=1/R$ to $q_{\text{max}}=2\pi/d$, with $d=0.5$ $\mu$m the focal diameter of the optical trap \cite{Betz2012}. (c) Low-frequency plateau of the power spectrum for adhesion-dominated fluctuations (\eq{eq_plateau}) as a function of the pressure on the membrane. } \label{fig_psd}
\end{center}
\end{figure}

Finally, the model in the previous section also provides dynamical information on membrane undulations. Specifically, the power spectral density of membrane fluctuations  can be shown to take the form \cite{Betz2009,Betz2012}
\begin{equation} \label{eq_psd}
S\left(\omega\right)=\frac{4\eta_c k_BT}{\pi}\int_{q_{\rm min}}^{q_{\rm max}}\frac{dq}{\left(4\eta_c\omega\right)^2+\left(\kappa q^3+\gamma q+\rho_{b,\rm eq}k/q\right)^2},
\end{equation}
where $q_{\rm min}$ and $q_{\rm max}$ are the cutoff values of the wave-vector $q$. In our model, either the perimeter of the cell, the correlation length of cortex undulations, or the radius of the pipette in the experimental setup proposed in \fig{fig_psd}a sets the large-wavelength cutoff, $q_{\rm min}\sim1/R$, and the short-wavelength cutoff is set by the spacing of the linkers: $q_{\rm max}=2\pi/\rho_0^{-1/2}$. In fluctuation spectroscopy experiments, the laser focal diameter sets the limitation for the latter \cite{Betz2009,Betz2012}. 

Membrane-cortex detachment induced by micropipette aspiration is a rather invasive procedure to assess the stability of the membrane-cortex cellular interface. An alternative approach could be to monitor membrane fluctuations for different aspiration pressures using fluctuation spectroscopy, as sketched in \fig{fig_psd}a. \fig{fig_psd}b shows the power spectrum density \eq{eq_psd} in the limit $\gamma\rightarrow 0$ both for bending-dominated and adhesion-dominated membrane fluctuations. The high-frequency limits were previously obtained: $S\left(\omega\right)\simeq k_BT/(6\left(2\kappa\eta_c^2\right)^{1/3})\omega^{-5/3}$ for $\lambda_cq_{\rm max}\gg1$, and $S\left(\omega\right)\simeq k_BTq_{\rm max}/(4\pi\eta_c)\omega^{-2}$ otherwise \cite{Helfer2001,Betz2009,Betz2012} (see more details in section 2.4 of the Supporting Material). However, our model gives an analytical expression for the full power spectrum in the adhesion-dominated regime ($q_{\text{max}}<\left[\rho_{b,\text{eq}} k/\kappa\right]^{1/4}$):
\begin{multline} \label{eq_fit}
\lim_{\kappa,\gamma\rightarrow 0}S\left(\omega\right)=\frac{k_BT}{4\pi\eta_c\omega^2}\left[q_{\rm max}-q_{\rm min}+\frac{\rho_{b,\rm eq}k}{4\eta_c\omega}\right.\\
\left.\times\left[\arctan\left(\frac{4\eta_c q_{\rm min}\omega}{\rho_{b,\rm eq}k}\right)-\arctan\left(\frac{4\eta_c q_{\rm max}\omega}{\rho_{b,\rm eq}k}\right)\right]\right].
\end{multline}
The density of membrane-cortex bonds $\rho_{b,\rm eq}$ can be extracted by fitting this expression to experimental measurements. In particular, if adhesion dominates membrane fluctuations, $\rho_{b,\rm eq}$ can be simply obtained from the plateau of the power spectrum at low frequencies:
\begin{equation} \label{eq_plateau}
\lim_{\omega\rightarrow 0}\lim_{\kappa,\gamma\rightarrow 0}S\left(\omega\right)=\frac{4\eta_c k_BT}{3\pi\left(\rho_{b,\rm eq}k\right)^2}\left(q_{\rm max}^3-q_{\rm min}^3\right).
\end{equation}
The value of this plateau is plotted in \fig{fig_psd}c as a function of the pressure on the membrane, $\Delta P$, which modifies the density of bound linkers. Experimentally, the pressure on the membrane can be varied either decreasing cortical tension by inhibiting myosin activity or via micropipette suction. Hence, we propose combined spectroscopy and micropipette experiments, as illustrated in \fig{fig_psd}a., to test the predictions in \fig{fig_psd} and estimate the density of membrane-cortex bonds. Note that the tip of the aspirated membrane is not flat, but is on average hemispherical with a radius of curvature matching the pipette radius. A rigorous analysis of the fluctuation spectrum should be done using spherical harmonics rather than Fourier transform. Furthermore, Eq.~\ref{eq_psd} does not account for the hard-wall repulsion introduced by the pipette walls. As discussed in \cite{Betz2012}, this introduces differences in the low frequency limit of the power spectrum. However, this should not affect the pressure dependence of the zero-frequency power spectrum shown in \fig{fig_psd}c. The correction to \eq{eq_fit} due to a finite average membrane curvature can be reduced by increasing the radius of the micropipette, or by tuning mysosin activity rather than using a micropipette to modify the average density of bond linkers.

The measurement of the density of membrane-cortex linkers from fluctuation spectroscopy is complementary to the quantitative determination of the cortical activity and adhesion energy from micropipette experiments, as discussed above. Indeed, data on fluctuation spectra of generic eukaryotic cells other that red blood cells are still lacking. Peukes and Betz have recently obtained such spectra in blebs during their growth stage, while the cortex is still reforming and, thus, weak \cite{Peukes2014}. However, information about the full cortex could only be extracted from experiments probing the fluctuations of strongly adhered membranes instead of blebs. Peukes and Betz analyze the fluctuation spectra as that of isolated membranes, with the effect of the cortex only incorporated into an effective tension of the membrane \cite{Peukes2014}. In contrast, our model accounts for the effect of the adhesion to the cortex via the kinetics of the linkers, thus providing a theoretical framework in which to consistently interpret fluctuation spectroscopy experiments on strongly adhered cell membranes.

As a final comment, it is worth stressing that in this paper we have only addressed passive fluctuations of thermal origin. In general, different active processes could potentially modify the presented scenario. Typically, active processes are quantitatively most pronounced at low frequencies. At high enough frequencies it has been shown that the role of active fluctuations can be incorporated through an increased effective temperature of the membrane \cite{Manneville1999,Manneville2001,Gov2003}. A detailed analysis of this point is beyond the scope of this work and is deferred to future work.

\section{Conclusions}
We have described a model for membrane-cortex adhesion that relates the unbinding pressure and adhesion energy measured in micropipette experiments to two cellular parameters, the membrane-cortex ligand density and the myosin-driven cortical activity. The validity of the model is qualitatively discussed although a complete set of experiments will be required for a complete validation. The proposed relationship between unbinding pressure and cortical activity provides a method to measure the cortical activity by means of micropipette aspiration experiments. Accounting for membrane undulations allows to relate the fluctuation spectrum of the membrane to the density of bound membrane-cortex bonds, thus providing a method for measuring this quantity in fluctuation spectroscopy experiments. Together, these experiments could give access to quantitative information about membrane-cortex adhesion in the framework of our model.

\section*{ACKNOWLEDGMENTS}
R.A. acknowledges support from Fundaci\'{o} ``la Caixa'', J.C. acknowledges financial support of the Ministerio de Econom\'{i}a y Competitividad under projects FIS2010-21924-C02-02 and FIS2013-41144-P, and the Generalitat de Catalunya under projects 2009 SGR 14 and 2009 SGR 878, and P.S. acknowledges support from the Human Frontier Science Program under the grant RGP0058/2011.

\section*{SUPPORTING MATERIAL}
Supporting Materials and Methods, six figures, and one table are available at http://\-www.\-biophysj\-.org/ biophysj/\-supplemental/\-S0006-3495(15)00226-X. References \cite{Guyon2001,Lin2005,DeGennes1982,Gov2004b,Safran1994,Ranft2012} appear in the Supporting Material.
\bibliography{Blebs}
\end{document}